# GP-DHT: A Dual-Head Transformer with Contrastive Learning for Predicting Gene Regulatory Relationships across Species from Single-Cell Data

Shuai Yan , Qingzhi Yu , Wengfeng Dai , Xiang Cheng[*]

School of Information Engineering, Jingdezhen Ceramics University, , Jingdezhen, 333403, Jiangxi, China

**Abstract**

Gene regulatory networks (GRNs) are essential for understanding cell-fate decisions and disease mechanisms, yet cross-species GRN inference from single-cell RNA-seq data remains challenging due to noise, sparsity, and cross-species distribution shifts. We propose GP-DHT (GenePair – DualHeadTransformer), a cross-species single-cell GRN inference framework that models genes and cells in a heterogeneous graph with multi-level expression relations and learns structured regulatory representations via multi-relational graph attention. A dual-head Transformer further captures local gene-pair regulatory dependencies and global cross-cell interaction patterns. To improve robustness under sparse and cross-species settings, GP-DHT introduces gene-pair–level supervised contrastive learning. Experiments on seven BEELINE benchmark datasets show consistent gains over representative baselines, improving AUROC and AUPRC by ~5%–7% on most datasets. GP-DHT also recovers known regulatory modules and helps distinguish conserved from species-specific regulations.

## 1 Introduction

Gene regulatory networks (GRNs) describe directed regulatory relationships between transcription factors (TFs) and their target genes and constitute a fundamental abstraction for understanding cell-fate decisions, developmental processes, and disease mechanisms. With the advent of single-cell technologies, GRN inference has progressively shifted from population-averaged expression analysis to cell-resolved modeling, enabling a more precise characterization of dynamic regulatory programs during cell-state transitions and lineage differentiation [1].

The rapid development of single-cell RNA sequencing (scRNA-seq) has provided unprecedented opportunities to investigate transcriptional heterogeneity at cellular resolution, thereby facilitating the identification of cell-type–specific and condition-dependent regulatory patterns [1, 2]. Consequently, GRN inference based on single-cell data has become an important research direction in computational biology and systems biology, leading to a wealth of methodological summaries and reviews[3, 4]. Recent advances have explored diverse modeling paradigms, including statistical frameworks [5, 6], machine learning approaches [7, 8], and deep neural architectures [7, 9], to better capture nonlinear dependencies and context-dependent regulation.

Despite this progress, scRNA-seq–based GRN inference remains challenging. Compared to bulk RNA-seq, single-cell data are characterized by severe sparsity, technical noise, and zero inflation, which obscure true regulatory signals and exacerbate indirect associations [10]. Regulatory relationships often vary across cell types, developmental stages, and species, further complicating inference. Traditional GRN methods relying on population-averaged expression struggle to capture nonlinear regulatory effects and are influenced by indirect associations, with RNA-level interactions further complicating inferences [11]. These challenges are particularly pronounced in cross-species settings, where expression distributions shift substantially and regulatory programs exhibit both conserved and species-specific components.

To address the limitations of traditional methods, deep learning approaches have been widely introduced to GRN inference in recent years to enhance representational capacity and modeling flexibility[12]. Neural network architectures can directly model the complex nonlinear dependencies between genes, forming an important branch of "deep learning for gene relationship inference"[13, 14]. Representative models, such as DeepDRIM, employ deep architectures to capture cell-type–specific regulatory patterns, while recent benchmark studies have systematically evaluated inference reliability across diverse datasets [13]. Nevertheless, many existing deep learning methods primarily focus on pairwise gene associations[5, 15] or local graph structures [16]and often lack mechanisms to explicitly incorporate global transcriptional context, limiting their interpretability and robustness.

Graph-based learning has emerged as a powerful paradigm for modeling structured biological data. Graph neural networks (GNNs) and attention mechanisms have been widely applied to single-cell analysis and GRN inference, enabling the integration of heterogeneous information and the modeling of complex dependencies [4]. Methods such as scGNN [17], GMFGRN [18] and HGAT-Link [19] leverage graph representations to improve regulatory edge prediction, while hypergraph and variational approaches further extend modeling capacity to higher-order regulatory modules [20]. In parallel, graph embedding and generative models for single-cell end-to-end representation learning are also being developed. For instance, a framework combining graph embedding and Gaussian mixture VAE has been used for end-to-end analysis of single-cell data, providing complementary insights into expression representation learning and structural modeling[21]. Furthermore, PMF-GRN uses variational inference to improve the robustness and stability of inference[22].

Beyond structural modeling, expression semantics enhancement and knowledge graph methods, like KEGNI, improve biological consistency and inference reliability, especially in low signal-to-noise scenarios[23]. Meanwhile, multi-omics integration has gradually become an important trend in GRN inference. For example, Graph-linked embedding has shown that integrating multi-omics can improve single-cell data integration and regulatory inference capabilities[24]. More general multi-omics integration frameworks have also been used for GRN prediction. In specific multimodal practices, SCENIC+ integrates ATAC-seq and TF motif data into the inference process to constrain candidate regulatory edges and reduce spurious regulation caused by co-expression[25]. Methods like SnapATAC provide foundational tools for the system analysis of single-cell ATAC-seq[26]. Similarly, large-scale functional genomics resources, such as ENCODE's cataloging of human/mouse regulatory elements, provide an important foundation for cross-species regulatory evidence integration and prior construction[27]. In addition, single-cell studies in specific biological systems have revealed the regulatory role of dynamic gene expression networks in B cell development and transformation, offering a reference for "network-level interpretability"[28].

While progress continues in single-species inference, the research focus has gradually expanded to the conservation and transferability of cross-species regulatory mechanisms. Early studies on cross-species network alignment have proposed statistical frameworks for cross-species network comparison[10]. Recently, cross-species single-cell representation learning methods have used shared latent spaces or foundational models to achieve expression alignment and knowledge transfer, such as GeneCompass, which constructs knowledge-driven cross-species foundational models to decipher universal regulatory mechanisms[29]. SATURN learns universal cell embeddings to reduce reliance on strict homologous mapping[30]. scPRINT, through pre-training on massive single-cell data, enhances the robustness and generalization of network predictions[31].

Despite significant progress, single-cell (especially cross-species) GRN inference still faces key challenges: First, simplifying continuous expression into traditional gene–gene graphs may lose expression gradient information, weakening the ability to capture regulatory cues[2]. Second, many models emphasize local structure or global dependencies, making it difficult to jointly capture local constraints and global context in a unified framework[3, 10]. Third, cross-species expression distribution shifts and species-specific regulatory differences significantly limit the generalization ability of models[2, 24]. Additionally, training data is often highly imbalanced, and negative sample construction and sampling strategies significantly affect learning outcomes. Research in the field of graph learning on "regional/global negative sampling principles" provides a methodological reference for designing more reasonable negative sampling mechanisms[32].

To address these challenges, we propose GP-DHT (GenePair–DualHeadTransformer), a unified cross-species single-cell GRN inference framework. GP-DHT integrates heterogeneous [1]graph learning, a dual-head Transformer architecture, and gene-pair-level contrastive learning strategies to jointly model expression gradient information, structured regulatory dependencies, and cross-species shared features. We systematically evaluated GP-DHT on seven single-cell datasets provided by the BEELINE benchmark framework[2]. The experimental results show that this method has superior prediction performance and stronger generalization ability in high-noise, highly sparse, and cross-species scenarios.

## 2 Methods

### 2.1 Datasets

We evaluated seven single-cell RNA-seq datasets from the BEELINE benchmark suite, covering different cell types. Following the BEELINE preprocessing protocol, we removed lowly expressed genes and selected the top 500 and 1,000 most variable genes, resulting in 14 datasets for analysis.

For negative sampling, traditional uniform random sampling selects an equal number of gene pairs from non-ground-truth regulatory relationships, which may introduce bias. To address this, we propose a co-occurrence filtering-based negative sampling strategy. We exclude known positive gene pairs and filter out pairs that are jointly highly expressed in ≥5% of cells. This ensures that negative samples are less likely to represent biologically meaningful relationships. Finally, we randomly sample negatives from the remaining gene pairs at a size 5–10× that of the positive samples, reducing the risk of contaminating the negative set. Sampling statistics are reported in Table 1.

Table 1. Summary statistics of the scRNA-seq transcriptomes and corresponding ChIP-seq–derived regulatory networks.

| Cell type | Number of cells | Number of TF genes | Genes (top 500/1,000) | Positive/negative sample ratio |
|---|---|---|---|---|
| hESC | 759 | 34 (34) | 815 (1260) | 4545/26361 (7084/40822) |
| hHEP | 426 | 30 (31) | 874 (1331) | 9939/18471 (15558/29299) |
| mDC | 384 | 20 (21) | 443 (684) | 756/15644 (1193/26527) |
| mESC | 422 | 88 (89) | 977 (1385) | 29613/68859 (42795/101296) |
| mHSC-E | 1072 | 29 (33) | 691 (1177) | 11557/8830 (21975/17724) |
| mHSC-GM | 890 | 22 (23) | 618 (1089) | 7364/6518 (14135/11878) |
| mHSC-L | 848 | 16 (16) | 525 (640) | 4398/4545 (5180/5876) |

**2.2 Data preprocessing and feature extraction**

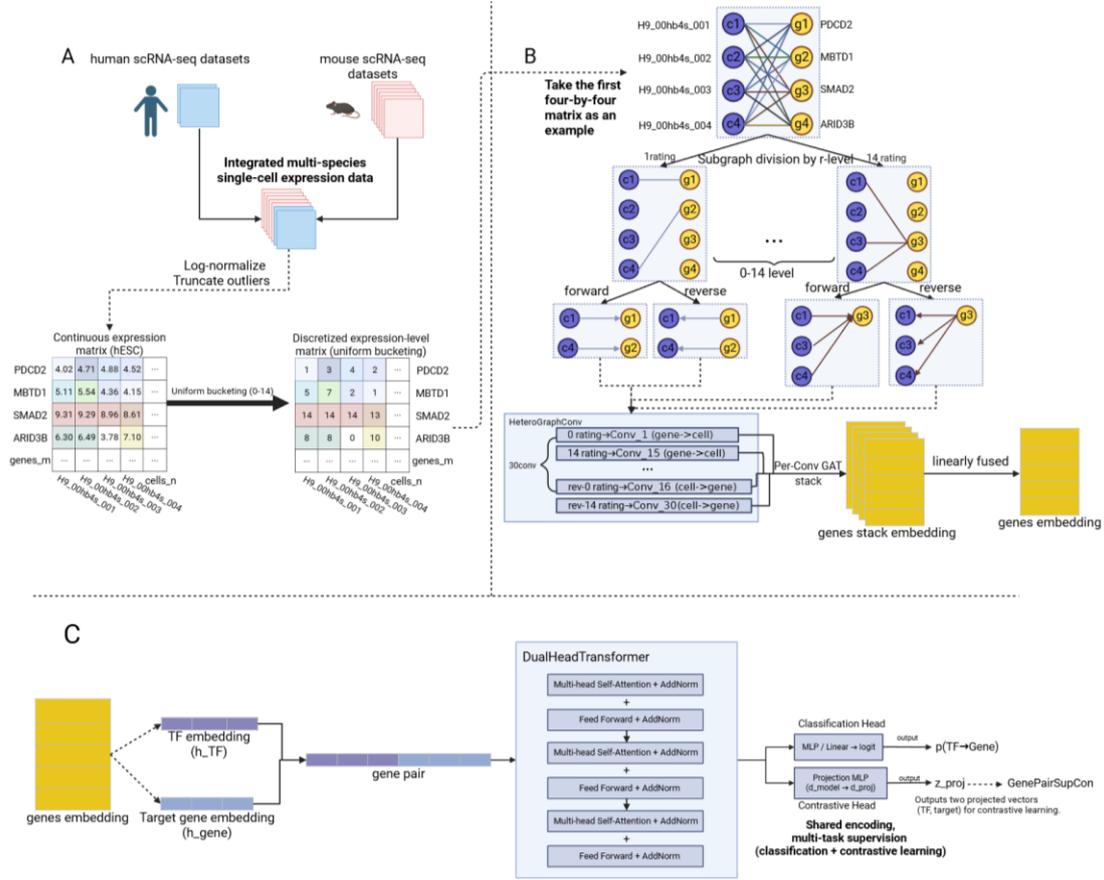

Figure 1. Overview of the GP-DHT framework.
(A) Data preprocessing and gene–cell heterogeneous graph construction from integrated human and mouse scRNA-seq data. Expression values are discretized into multi-level relations via binning to form a gene–cell graph.
(B) Heterogeneous graph encoding with relation-specific GAT message passing on expression-level subgraphs, followed by multi-relation fusion to obtain gene embeddings.
(C) Gene-pair modeling and prediction. TF and target-gene embeddings are paired and fed into a dual-head Transformer to capture local gene-pair dependencies and global contextual patterns. A classification head predicts regulatory relationships, while a supervised contrastive objective (GenePairSupCon) provides auxiliary training signals.

### 2.2.1 Data processing

Preprocessing is critical for ensuring the accuracy and stability of single-cell transcriptomic analyses. We perform several steps—gene filtering, log transformation, value clipping, and discretization—to enhance data quality. First, lowly expressed genes are removed as they contribute minimally to the model and introduce noise.

We then apply log transformation to the remaining gene expression values to compress the dynamic range and mitigate the impact of extreme values. Zero expressions are preserved to maintain sparsity. The log transformation is defined as:

$$x'_{ij} = \begin{cases} \ln(x_{ij}), & x_{ij} > 0 \\ 0, & x_{ij} = 0 \end{cases} \quad i \in 1,2,\dots,g; \quad j \in 1,2,\dots,c \quad (1)$$

Next, extreme values are clipped. For each gene, we calculate the mean $\mu_i$ and standard deviation $\sigma_i$, defining the effective range $[\mu_i - 2\sigma_i, \mu_i + 2\sigma_i]$. Values outside this range are clipped, reducing the influence of outliers. This range is computed as:

$$\mu = \frac{1}{n}\sum_{i=1}^{n}\log(x_i), \quad \sigma = \sqrt{\frac{1}{n-1}\sum_{i=1}^{n}(\log(x_i)-\mu)^2} \quad (2)$$

After clipping, we discretize expression values by mapping them to discrete levels based on predefined intervals. Out-of-range values are assigned to the lowest or highest level. The discretization process is formulated as:

$$x''_{ij} = \begin{cases} k, & x'_{ij} \in \text{Range}_{ik} \\ 1, & 0 \le x'_{ij} < \mu_i - 2\sigma_i \\ n, & x'_{ij} > \mu_i + 2\sigma_i \end{cases} \quad (3)$$

These preprocessing steps reduce noise, suppress extreme values, and transform the data into a suitable form for downstream modeling. The discretization process further helps extract meaningful features for regulatory network inference.

### 2.2.2 Expression-level binning strategy

We investigate two binning strategies for discretizing single-cell gene expression data: **Uniform Bucketing** and **Cumulative Bucketing**, each with different trade-offs in preserving continuity and producing discrete levels, affecting GRN inference.

**Uniform Bucketing:**

In this strategy, we first apply a log transformation to the gene expression values $x_g$, then partition the resulting range into $K$ levels. The procedure is as follows:

**Log transformation**:
$$\ell_g = \log(x_g) \quad (4)$$

**Level Boundaries**: The mean $\mu$, standard deviation $\sigma$, and min/max values of the log-transformed expression values are computed to define the level boundaries:
$$[L, U] = [\max(\mu - 2\sigma, t_{\min}), \min(\mu + 2\sigma, t_{\max})] \quad (5)$$

**Level Assignment**: Expression values are assigned to discrete levels based on their position in the defined range:
$$b_{\mathrm{uni}}(x_g) = \begin{cases} 0, & \text{if } x_g = 0 \\ \min\{K-1, \max(0, \lfloor(\log(x_g) - L)/w\rfloor)\} + 1, & \text{if } x_g > 0 \end{cases} \quad (6)$$

**RPKM normalization**:
$$r_{\mathrm{uni}}(x_g) = \begin{cases} \dfrac{\log(x_g) - L}{w} & \text{if } x_g > 0 \\ 9999 & \text{if } x_g = 0 \end{cases} \quad (7)$$

**Cumulative Bucketing:**

This strategy uses a pyramid-style binning, where each gene expression value is assigned to multiple levels, including all levels up to its own. The key steps are:

**Level Calculation**: We first compute the level for each gene using uniform bucketing. Then, the effective number of levels is determined by:
$$L^\star = b_{\mathrm{uni}}(x) \quad (8)$$

**Level Expansion**: Each gene–cell pair is assigned to all levels up to $L^\star$, creating additional edges in the graph:
$$E = \bigcup_{l=1}^{L^\star} (g \to c), (c \to g) \quad (9)$$

**RPKM Normalization**: Similar to uniform bucketing, RPKM values are computed for each gene across all levels:
$$r_{\mathrm{cumulative}}(x_g) = \begin{cases} \dfrac{\log(x_g) - L}{w} & \text{if } x_g > 0 \\ 9999 & \text{if } x_g = 0 \end{cases} \quad (10)$$

**Computational Complexity**

**Uniform Bucketing**: Each non-zero entry leads to one edge, resulting in $2M$ edges, where $M$ is the number of non-zero elements.

**Cumulative Bucketing**: Each non-zero entry connects to multiple level relations, resulting in $O(KM)$ edges, where $K$ is the number of bins.

Both strategies increase graph complexity and memory consumption with larger $K$, but provide richer relational information to improve prediction accuracy.

### 2.2.3 Gene–cell representation learning: embedding model construction

We propose a heterogeneous-graph representation learning framework to model regulatory relationships between genes and cells, treating genes and cells as node types, with gene–cell pairs forming edges. Edge types are defined by discretized expression levels, with reverse edges added to enhance information propagation.

This design captures multi-dimensional regulatory relationships. Genes and cells are encoded using one-hot vectors. During training, node features are propagated and aggregated using graph neural networks (GNNs) and attention mechanisms to learn embeddings.

To capture differential interactions at varying expression levels, we introduce a multi-relational graph attention mechanism, creating independent message-passing pathways for each edge type. Node features are projected into relation-specific spaces, and attention mechanisms compute the weights for updating target nodes, enabling accurate modeling of gene–cell regulatory patterns.

For each edge type $r$, representing a specific expression level and direction, node pairs $(u, v)$ denote the source and target nodes, with features projected using edge-specific transformations.

$$h_u^{(r)} = W_r h_u, \quad h_v^{(r)} = W_k h_v \quad (11)$$

The input features of nodes $u$ and $v$ are denoted as $\mathbf{h}_u$ and $\mathbf{h}_v$, respectively, and $\mathbf{W}_r$ and $\mathbf{W}_k$ are relation-specific learnable parameters.

Next, the model computes attention weights within the relation-specific subspace by normalizing the features and calculating element-wise similarity to obtain the unnormalized attention response. A temperature parameter $\tau$ is introduced for smoothing.

$$\alpha^{(r)} = \mathrm{Softmax}\left(\frac{|\mathrm{Norm}(\tilde{h}_v^{(r)}) \odot \mathrm{Norm}(\tilde{h}_v^{(r)})|}{\tau}\right) \quad (12)$$

These attention weights characterize, under a specific relation type, the importance of different feature dimensions for updating the target-node representation.

During message passing, the model first aggregates source-node features within the neighborhood of the same edge type (using a max-pooling operation). It then modulates the aggregated message with the relation-specific attention weights, and applies a nonlinear activation function to obtain the updated node representation:

$$h_v^{(r)'} = \sigma\big(\text{AGG}_{u \in \mathcal{N}_r(v)}(\tilde{h}_u^{(r)}) \odot \alpha^{(r)}\big) \quad (13)$$

The concatenated high-dimensional features are then projected back to the target dimension $d_{\text{out}}$ through a linear layer, yielding the final representation of node v:

$$h_v = W_o \|_{r \in \mathcal{R}} h_v^{(r)'} + b \quad (14)$$

The heterogeneous graph encoding is performed in parallel across all gene and cell nodes. After one layer of message passing, the model generates gene and cell embedding matrices $G \in \mathbb{R}^{N_g \times d}$ and $C \in \mathbb{R}^{N_c \times d}$, respectively. These embeddings capture the multi-level expression features of genes across cells and the regulatory information received by cells, providing high-quality input for the dual-head Transformer decoder, which is used for classification and supervised contrastive learning.

## 2.3 Gene regulatory prediction via a dual-head Transformer

After obtaining embedding representations for genes and cells, a key step in single-cell GRN inference is how to effectively map these representations to reliable gene regulatory relationships. To this end, we design a dual-head Transformer – based regulatory prediction module, which explicitly distinguishes and jointly models (i) local regulatory signals at the gene-pair level and (ii) global transcriptional-context dependencies driven by cell states. This design is motivated by the biological nature of transcriptional regulation: true regulatory relationships depend not only on direct interactions between transcription factors and target genes, but are also jointly modulated by the broader transcriptional environment, including cell type, developmental stage, and species background.

### 2.3.1 Design rationale: advantages of the dual-head Transformer

In GRN inference, accurately modeling regulatory relationships between gene pairs is crucial. To capture both local and global signals, we propose a dual-head Transformer architecture. The dual-head design processes information through two independent heads: one for regulatory prediction and the other for contrastive learning. This design improves performance and robustness by optimizing for both objectives, especially in cross-species and multi–cell-state settings.

The regulatory prediction head focuses on predicting relationships at the gene-pair level, using the Transformer to capture dependencies between genes. Its output is a prediction of whether a regulatory relationship exists between a given gene pair.

$$H_{\text{class}} = \text{ClassHead}(X_{\text{genepair}}) \quad (15)$$

Contrastive learning head. This head incorporates a contrastive learning mechanism to enhance model robustness. It learns to map related gene pairs to nearby regions in the embedding space while pushing unrelated gene pairs farther apart. This objective helps the model better adapt to sparse data and cross-species settings.

$$H_{\text{contrast}} = \text{ContrastHead}(X_{\text{genepair}}) \quad (16)$$

The two heads serve different purposes at different stages of the model. The output of the regulatory prediction head (also referred to as the classification head) is used to produce the final prediction of regulatory relationships between gene pairs. In contrast, the contrastive learning head is used to compute the contrastive loss, which refines gene-pair embeddings during training and thereby improves the robustness of the model.

### 2.3.2 Transformer encoder architecture

We employ a Transformer encoder with three layers, each consisting of multi-head self-attention, a position-wise feed-forward network (FFN), and layer normalization. This architecture enables the model to capture complex nonlinear dependencies between gene pairs, discovering latent associations without relying on a predefined graph topology.

The gene-pair embedding matrix $X_{\text{genepair}}$ is fed into the encoder. In each layer, self-attention models interactions among input features, enhancing the expressiveness of the learned representations. Multi-head self-attention computes

the weighted sum of the value matrix $V$ using query $Q$ and key $K$ matrices, as shown below:

$$\text{Attention}(Q, K, V) = \text{softmax}\left(\frac{Q^\top K}{\sqrt{d_k}}\right) V \quad (17)$$

Next, the feed-forward network (FFN) applies a GELU activation function:

$$\text{FFN}(x) = \text{GELU}(xW_1 + b_1)W_2 + b_2 \quad (18)$$

The output of each layer is normalized using layer normalization:

$$\text{LayerNorm}(x) = \gamma \cdot \frac{x - \mathbb{E}[x]}{\text{Var}[x]} + \epsilon + \beta \quad (19)$$

Finally, the encoder output $h$ is mapped to a 256-dimensional representation using a fully connected layer with GELU activation and layer normalization:

$$f = \text{LayerNorm}(\text{GELU}(hW + b)) \quad (20)$$

This design effectively captures complex dependencies between gene pairs.

### 2.3.3 Robust representation constraints via contrastive learning

GRN inference from single-cell transcriptomic data faces challenges such as data sparsity, noise, and cross-species expression shifts. Conventional supervised learning methods struggle in these settings, often leading to overfitting and limited generalization. To address this, we introduce contrastive learning to improve model robustness and generalization.

Contrastive learning optimizes gene-pair embeddings by maximizing the similarity of positive pairs (gene pairs with regulatory relationships) and minimizing that of negative pairs (non-regulatory gene pairs). This approach helps the model better capture regulatory signals between genes under sparse and cross-species data conditions.

Within the dual-head Transformer, we add an independent contrastive learning head, which optimizes gene-pair embeddings by computing pairwise similarities. The contrastive loss, based on cosine similarity, encourages higher similarity for positive pairs and lower similarity for negative pairs. The loss is defined as:

$$L_{\text{SupCon}}(h_i, h_j) = -\log \frac{exp(\text{sim}(h_i, h_j)/\tau)}{\sum_{k \neq i} exp(\text{sim}(h_i, h_k)/\tau)} \quad (21)$$

where $\text{sim}(h_i, h_j)$ is the cosine similarity between gene-pair embeddings, and $\tau$ is a temperature parameter controlling the sharpness of the similarity distribution.

### 2.3.4 Joint optimization of classification and contrastive losses

To improve robustness and generalization under sparse conditions, we introduce a weighted loss function that combines classification and contrastive losses. This helps the model predict regulatory relationships while refining gene-pair representations using contrastive learning.

We also apply data augmentation by injecting Gaussian noise into the training data, increasing diversity and helping the model learn stable features. Specifically, Gaussian noise is added to the input features $X$, generating augmented data with controlled noise strength.

The model is trained using binary cross-entropy loss for regulatory predictions and contrastive loss for robustness. The total loss is:

$$L_{\text{total}} = L_{\text{BCE}} + \lambda \cdot L_{\text{SupCon}} \quad (22)$$

where $L_{\text{BCE}}$ is the binary cross-entropy loss, $L_{\text{SupCon}}$ is the supervised contrastive loss, and $\lambda$ is a weighting factor.

We compute losses on both the original and augmented data and average them to fully utilize augmented samples during training.

The final training objective combines both losses:

$$L_{\text{total}} = 0.6 \cdot L_{\text{BCE, total}} + 0.4 \cdot L_{\text{SupCon, total}} \quad (23)$$

Hyperparameters, including learning rate and weight decay, were tuned on a validation set, with an embedding dimension of 256 yielding optimal performance.

## 3 Experiments and Results

To evaluate GP-DHT's performance on single-cell and cross-species GRN inference tasks, we use AUROC and AUPRC as primary metrics. AUPRC is more sensitive to class imbalance, while AUROC measures overall performance. Together, they provide a comprehensive assessment of the precision-recall trade-off.

We use a three-fold cross-validation protocol, with 33.3% of samples for testing and the rest for training and validation. Instead of averaging across folds, we aggregate all test samples for a more stable evaluation, reducing the impact of small-sample fluctuations.

For model optimization, we use the Adam optimizer with a batch size of 512. Training stops after 200 iterations or if parameters are not updated for 10 consecutive steps. We compare GP-

DHT with several baseline methods: HGAT-Link[17, 19] (combining heterogeneous graph attention networks with a Transformer to improve GRN inference), GMFGRN[18] (leveraging heterogeneous information and heterogeneous-graph embeddings to reduce false positives), IGEGRNS [20] (learning embeddings via GraphSAGE and producing final predictions after three convolutional layers), GENELink [33] (integrating regulatory information and learning robust, general gene representations under supervision of a graph attention network), and GATCL[34] (performing GRN inference by learning graph attention using convolutional layers instead of weight matrices).

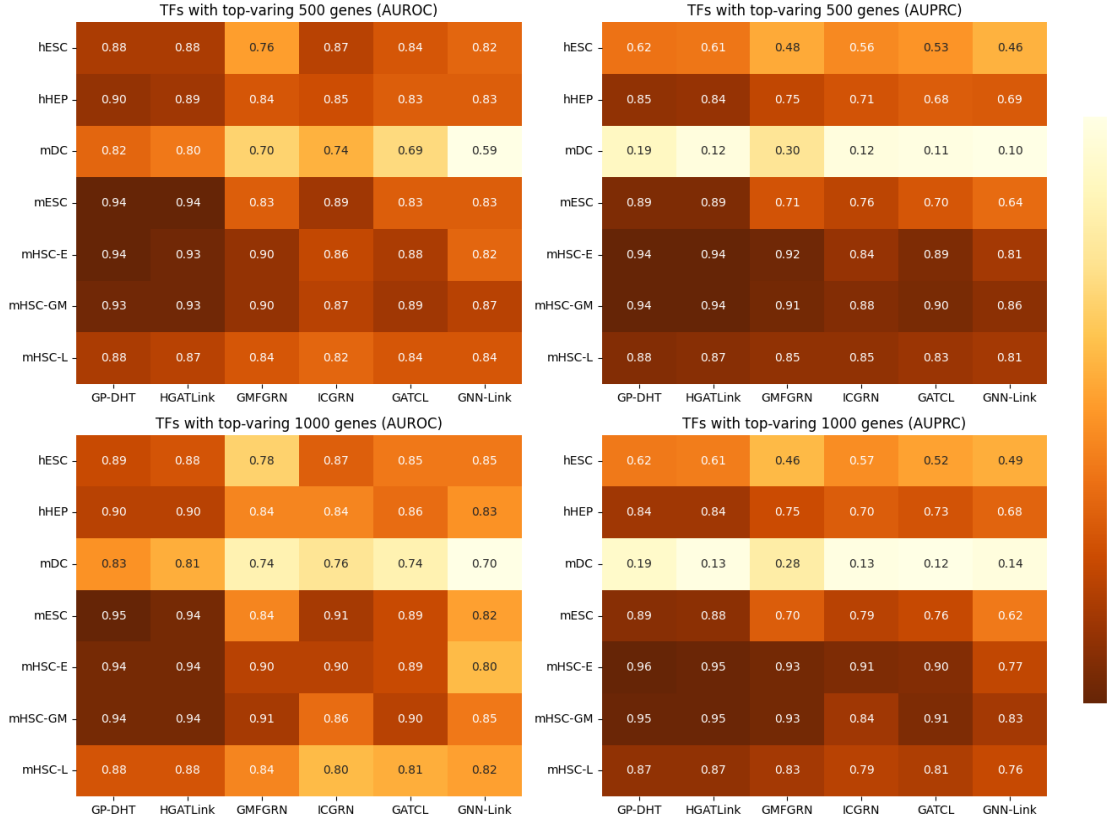

Figure 2. Performance comparison between GP-DHT and five baseline methods (HGATLink, GMFGRN, IGEGRN, GATCL, and GNN-Link) across seven single-cell datasets. Results are reported using AUROC (left) and AUPRC (right) with the top 500 (top row) and top 1,000 (bottom row) most variable genes.

### 3.2 Effect of the binning strategy

We evaluated two binning strategies, Uniform Bucketing and Cumulative Bucketing. All experiments were conducted under the same hyperparameter configuration, and model performance was assessed using AUC and AUPR. The results show that Uniform Bucketing achieves higher performance than Cumulative Bucketing; therefore, we adopt Uniform Bucketing as the default binning scheme in the final model.

**Table 2.** Comparison of different binning strategies for gene expression discretization.

| Dataset | Uniform Bucketing | | Cumulative Bucketing | |
|---|---|---|---|---|
| | AUC | AP | AUC | AP |
| hESC1000 | 0.89 | 0.62 | 0.86 | 0.54 |
| hHEP1000 | 0.90 | 0.84 | 0.85 | 0.72 |
| mDC1000 | 0.83 | 0.19 | 0.78 | 0.12 |
| mESC1000 | 0.95 | 0.89 | 0.93 | 0.86 |
| mHSC-E1000 | 0.94 | 0.96 | 0.92 | 0.93 |
| mHSC-GM1000 | 0.94 | 0.95 | 0.89 | 0.90 |
| mHSC-L1000 | 0.88 | 0.86 | 0.86 | 0.85 |
| hESC500 | 0.88 | 0.62 | 0.86 | 0.55 |
| hHEP500 | 0.90 | 0.85 | 0.85 | 0.74 |
| mDC500 | 0.82 | 0.19 | 0.79 | 0.12 |
| mESC500 | 0.94 | 0.89 | 0.93 | 0.87 |
| mHSC-E500 | 0.94 | 0.94 | 0.91 | 0.93 |
| mHSC-GM500 | 0.93 | 0.94 | 0.92 | 0.92 |

As shown in Table 2, the Uniform Bucketing strategy achieves higher AUC and AUPRC on most datasets. This is mainly because cumulative bucketing introduces a large number of overlapping edges, which amplifies redundant signals from highly expressed genes and causes local noise to be aggregated multiple times. In contrast, uniform bucketing preserves only mutually exclusive levels, reducing the number of edges while retaining discretized information about expression gradients, thereby improving the representation efficiency of the heterogeneous graph.

The results in Table 2 indicate that uniform bucketing yields better AUC and AUPRC on the majority of datasets. By comparison, although cumulative bucketing introduces richer relational information at the expression level, it also substantially increases edge redundancy and magnifies the noise accumulation effect associated with highly expressed genes.

By maintaining the exclusivity of expression levels, uniform bucketing effectively preserves expression-gradient information while controlling graph-structure complexity, achieving a more favorable trade-off between representational efficiency and predictive performance.

### 3.3 Visualization analysis

To validate the predictive performance and biological interpretability of GP-DHT, we generate three types of visualizations for each dataset: Precision–Recall (PR) curves, Top-K regulatory network graphs, and TF–target gene (TF–TG) heatmaps. These visualizations offer insights into model performance, network structure, and biological relevance.

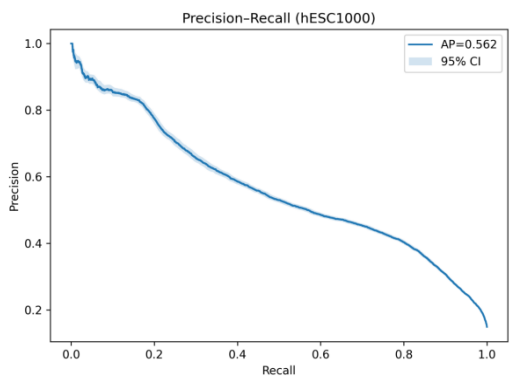

Figure 3. Precision–recall (PR) curves illustrating the performance of GP-DHT across different single-cell datasets.

First, the PR curves (Figure 3) illustrate how precision and recall vary under different decision thresholds. Across datasets, the average precision (AUPRC) consistently exceeds the random-classification baseline, with relatively narrow 95% confidence intervals, indicating that GP-DHT maintains stable discriminative capability across multiple cell types and species. Notably, on stem-cell datasets such as hESC1000 and mESC1000, the AUPRC exceeds 0.55, demonstrating the model's effectiveness in identifying true regulatory relationships.

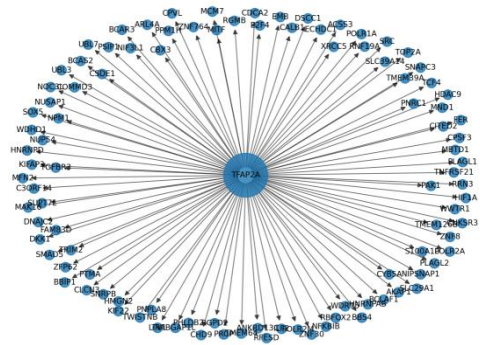

Figure 4. Visualization of the top-100 predicted regulatory edges in the inferred gene regulatory networks across different single-cell datasets.

Second, the Top-K regulatory network graphs (Figure 4) visualize the top 100 transcriptional regulatory edges with the highest predicted scores. In these graphs, nodes represent genes, and edges indicate regulatory direction, with arrows pointing from transcription factors to target genes. The predicted networks in most datasets show a "core-regulator–dominated" structure. For example, in the human embryonic stem cell dataset (hESC1000), TFAP2A emerges as a central hub connected to many downstream targets, suggesting that the model can identify key transcription factor–centered regulatory modules involved in development and differentiation.

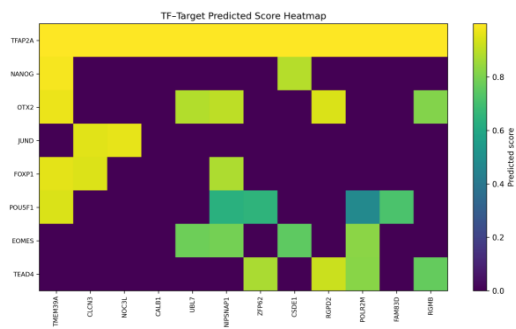

Figure 5. Heatmap of predicted regulatory scores between transcription factors and target genes.

Finally, the TF–target heatmap (Figure 5) presents predicted scores between transcription factors and target genes in a two-dimensional matrix. Yellow regions indicate strong regulatory relationships, while purple regions indicate weak or absent ones. Core factors such as TFAP2A, NANOG, and POU5F1 show consistently high predicted values across multiple targets, aligning with their known roles in pluripotency maintenance. These observations suggest that GP-DHT not only achieves superior quantitative performance but also captures biologically meaningful regulatory signatures.

In summary, GP-DHT demonstrates high accuracy and robustness for cross-species single-cell gene regulatory relationship prediction, while also showing strong interpretability in terms of network topology and biological relevance. This provides a reliable foundation for downstream functional validation and pathway analyses.

## 4 Discussion and Conclusion

This study presents GP-DHT, a unified framework for cross-species gene regulatory network (GRN) inference from single-cell RNA sequencing (scRNA-seq) data, designed to address key challenges arising from expression noise, data sparsity, and cross-species distribution shifts. By integrating heterogeneous graph modeling, a dual-head Transformer architecture, and gene-pair–level supervised contrastive learning, GP-DHT provides an effective solution for modeling complex transcriptional regulatory relationships across diverse biological contexts.

Extensive evaluations on seven benchmark datasets from the BEELINE framework demonstrate that GP-DHT consistently outperforms state-of-the-art GRN inference methods, including HGATLink, GMFGRN, and GNN-Link, achieving improvements of approximately 5%–7% in both AUROC and AUPRC across most datasets. These performance gains indicate that explicitly modeling expression-gradient information and gene–cell heterogeneity substantially enhances regulatory signal extraction under noisy and sparse conditions. In particular, the heterogeneous graph construction enables GP-DHT to preserve multi-level expression relations that are often lost in conventional gene–gene graph formulations.

A central contribution of GP-DHT lies in its dual-head Transformer design, which explicitly separates local gene-pair regulatory signals from global transcriptional-context dependencies. Unlike many existing GNN-based GRN inference methods that primarily emphasize either local graph topology or global embedding similarity, GP-DHT decomposes regulatory evidence into two complementary components: (i) local dependencies between transcription factors and target genes, and (ii) global contextual patterns driven by shared cellular states and species-specific backgrounds. This architectural distinction not only improves predictive performance but also provides a more interpretable framework for understanding how regulatory relationships are supported by both direct gene-pair interactions and broader transcriptional environments.

Ablation studies further validate the necessity of the proposed architectural components and design choices. In particular, the comparison between uniform and cumulative bucketing strategies highlights the importance of controlling edge redundancy in heterogeneous graphs. Uniform bucketing achieves a more favorable balance between representational efficiency and noise suppression by maintaining mutually exclusive expression levels, thereby improving downstream prediction accuracy. These findings suggest that careful discretization and relation design are critical for effective graph-based modeling of single-cell transcriptomic data.

Beyond quantitative performance, GP-DHT exhibits strong biological interpretability through multiple visualization analyses. Precision–recall curves indicate stable discriminative capability across cell types and species, while Top-K regulatory subnetworks and TF–target gene heatmaps reveal coherent regulatory structures centered around well-known transcription factors. Importantly, the interpretability offered by GP-DHT operates at the network and module level, enabling the identification of core regulators and conserved regulatory patterns rather than attempting residue-level or causal mechanistic explanations. This level of interpretability is well aligned with the current goals of GRN inference from observational single-cell data.

Despite its promising performance, GP-DHT has several limitations. First, the framework relies on static ChIP-seq–derived regulatory priors, which may not fully capture dynamic or context-specific regulatory changes. Incorporating time-resolved data, perturbation experiments, or causal intervention signals represents an important direction for future research. Second, the computational cost associated with heterogeneous message passing and Transformer-based modeling may limit scalability to extremely large graphs. Future work may explore sparse attention mechanisms, subgraph sampling strategies, and parameter-sharing schemes to improve computational efficiency without sacrificing predictive accuracy. Additionally, extending GP-DHT to integrate multimodal data sources, such as chromatin ac-cessibility and motif information, may further enhance cross-species generalization and biological relevance.

In conclusion, GP-DHT provides a robust and interpretable framework for cross-species GRN inference from single-cell data by unifying heterogeneous graph learning, Transformer-based modeling, and contrastive optimization. By explicitly modeling both local regulatory interactions and global transcriptional context, GP-DHT advances the methodological foundation for gene regulatory analysis and offers a flexible platform for future developments in dynamic, multimodal, and causal regulatory network modeling.